\documentclass[prd,
,superscriptaddress,showpacs,nofootinbib,%
tightenlines]{revtex4}
\usepackage{graphicx}
\usepackage{amssymb,amsmath}
\usepackage{ulem}
\newcommand{\beq}{\begin{equation}}
\newcommand{\eeq}{\end{equation}}
\newcommand{\beqa}{\begin{eqnarray}}
\newcommand{\eeqa}{\end{eqnarray}}

\newcommand{\ben}{\begin{displaymath}}
\newcommand{\een}{\end{displaymath}}
\newcommand{\be}{\begin{equation}}
\newcommand{\ee}{\end{equation}}
\newcommand{\bea}{\begin{eqnarray}}
\newcommand{\eea}{\end{eqnarray}}

\usepackage{color}

\begin{document}
\title{
How to renormalize integral equations with singular potentials \\ in effective field theory}
\author{E.~Epelbaum}
\affiliation
{Ruhr University Bochum, Faculty of Physics and Astronomy,
Institute for Theoretical Physics II, D-44870 Bochum, Germany}
\author{A.~M.~Gasparyan}
\affiliation
{Ruhr University Bochum, Faculty of Physics and Astronomy,
Institute for Theoretical Physics II, D-44870 Bochum, Germany}
\affiliation{NRC "Kurchatov Institute" - ITEP, B. Cheremushkinskaya 25, 117218 Moscow, Russia}
\author{J.~Gegelia}
\affiliation
{Ruhr University Bochum, Faculty of Physics and Astronomy,
Institute for Theoretical Physics II, D-44870 Bochum, Germany}
\affiliation{Tbilisi State  University,  0186 Tbilisi, Georgia}
\author{Ulf-G.~Mei\ss ner}
\affiliation{Helmholtz Institut f\"ur Strahlen- und Kernphysik and Bethe
   Center for Theoretical Physics, Universit\"at Bonn, D-53115 Bonn, Germany}
\affiliation{Institute for Advanced Simulation, Institut f\"ur Kernphysik
   and J\"ulich Center for Hadron Physics, Forschungszentrum J\"ulich, D-52425 J\"ulich,
   Germany}
\affiliation{Tbilisi State  University,  0186 Tbilisi, Georgia}
\author{X.-L.~Ren}
\affiliation
{Ruhr University Bochum, Faculty of Physics and Astronomy,
Institute for Theoretical Physics II, D-44870 Bochum, Germany}

\date{Jan. 17, 2020}
\begin{abstract}
We briefly review general concepts of renormalization in quantum field
theory and discuss their application
to solutions of integral equations with singular potentials in the few-nucleon sector of the low-energy effective field theory of QCD.
We also describe a particular subtractive renormalization scheme and consider a specific application to a toy-model with
a singular potential serving as its effective field theoretical leading-order approximation. 
\end{abstract}

\pacs{13.40.Gp,11.10.Gh,12.39.Fe,13.75.Cs}

\maketitle

\section{\label{introduction}Introduction}

The issue of renormalization is  a subject of long-standing controversy in the few-nucleon sector of the low-energy effective
field theory (EFT) of the strong interactions, pioneered in Ref.~\cite{Weinberg:rz}.
Recent reviews and references addressing this problem can be found,  e.g., in
Refs.~\cite{Bedaque:2002mn,Epelbaum:2005pn,Epelbaum:2008ga,Machleidt:2011zz,Epelbaum:2012vx,Birse:2009my,Valderrama:2016koj,Hammer:2019poc,Machleidt:2020vzm}.
The Lagrangian of any  consistent  EFT includes all local interactions
allowed by the underlying symmetries. Therefore, all ultraviolet (UV) divergences appearing
in physical quantities are in principle canceled by the corresponding counterterms \cite{Weinberg:mt}.
However, the practical implementation of such a quantum field theoretical renormalization is a highly non-trivial problem in
the few-body sector of the low-energy EFT of the strong interactions. The difficulty is caused by the fact that 
due to the
large absolute values of the scattering lengths,
an infinite number of {\it renormalized} diagrams has to be summed up at any order.
To perform this summation one defines an effective potential as a sum of irreducible diagrams contributing to the process
under consideration and then obtains the scattering amplitude by solving the corresponding integral equation \cite{Weinberg:rz}. 
In practice, solving integral equations corresponds to summing up {\it regularized} but {\it not renormalized} diagrams.
Regularized non-perturbative expressions obtained by  solving such integral equations, when expanded in powers of $\hbar$
(corresponding to the loop expansion), reproduce the regularized perturbative diagrams.
While non-perturbative solutions to integral equations may indeed contain pieces which yield vanishing contributions to the perturbative series,
the integral equations in the low-energy EFT are used solely as
tools for performing partial resummations of diagrams. Therefore,
properly renormalized non-perturbative expressions in a
self-consistent EFT must reproduce the renormalized perturbative series, when 
expanded in $\hbar$. 

It is well known how to renormalize Feynman diagrams in quantum field theories. However, it turns out to be a challanging
problem to properly renormalize the non-perturbative solutions to
integral equations since this usually requires
taking into account contributions of an infinite number of counterterms.
The problem  is actually caused by the non-renormalizability of EFT in the traditional sense, manifested by the presence of
singular potentials in the few-body sector.  
 
The low-energy EFT of the strong interactions yields a systematic expansion of the effective potentials in terms of small
masses and momenta, where small is taken relative to the typical hard (breakdown) scale of about 500~MeV. However, these expressions are
to be trusted only at low energies. Translated into  coordinate space, the EFT yields an expansion of the potential 
applicable at large distances, while the (strongly scheme-dependent)
short-range part is determined by the desired
resolution. Accordingly, extrapolations of the long-range part of the effective
potential to short distances do not root in the underlying QCD. 
Moreover, QCD features only shallow bound states in multi-nucleon
systems thus indicating that the singular van der Waals-like behavior of
one- and multi-pion exchange potentials is not a valid approximation of
the nuclear interaction at short distances.

Reading the literature in the few-body sector of low-energy chiral EFT  
and references therein, one often gets the impression that there is no 
well-established concept of renormalization for non-perturbative problems in quantum field theories, and that we are currently
seeking for such a self-consistent definition.
In this quest for the solution of a technically challenging problem, fundamental
concepts that are well established and should be well-known to the practitioners are often 
ignored or even disputed
in the nuclear physics community. Our understanding is that the very definition of the
EFT and
renormalization as it is understood in quantum field theory (QFT) specifies
uniquely what needs to be done. The implementation of 
a proper quantum field theoretical renormalization is, however, a rather challenging problem  
and subject to further investigations.

In our recent work on the renormalization in the few-body sector of the EFT of the strong interactions
\cite{Epelbaum:2018zli} we have put the main emphasis  on discussing  details of common misconceptions,
i.e. what should {\it not} be done 
when renormalizing the solutions to integral equations with singular EFT potentials.  
To summarize, the origin of the differences can be traced back to the application or enforcement of quantum mechanical renormalization to a quantum field theoretical problem.
This (in our opinion, incorrect) point of view is described in
detail in a recent review article \cite{vanKolck:2020llt}, see also
references therein. 
It amounts to handling singular potentials emerging in EFT contexts as
if they were quantum mechanical potentials valid at all distances rather than 
large-distance approximations to infinite series. We refer the reader to
Ref.~\cite{Epelbaum:2018zli} for more details.

In this paper we again revisit the issue of renormalization in few-body sector of chiral EFT,
but focus on {\it what should be done} to achieve the proper
renormalization in EFT with singular potentials. 
 
\medskip

Our paper is organized as follows.  In section~\ref{ren} we discuss the definition and conceptual issues of renormalization
in QFTs. In section~\ref{contactintpotential} we consider a particular QFT renormalization scheme
which can be applied to the solutions of integral equations with singular potentials. 
An application of the introduced renormalization scheme to a singular potential for a specific toy model is given in
section~\ref{OPE}. We summarize our work  in Sec.~\ref{summary}.

\section{Renormalization}
\label{ren}

Renormalization is a very natural and familiar procedure which is actually carried out in most theories. 
Considering a theory depending on the  parameters $g_i$, with $ i=1,...,N$,
one fixes these parameters from $N$ experiments and makes predictions for the results of other experiments.
The standard example is QED, where the electron charge and mass need to be determined from experiment and
then e.g. the anomalous magentic moment of the electron can be predicted to high precision.
In mathematical language this means that we calculate $N$ ``reference''  physical quantities
$\sigma_i(E)=f_i(\hbar, g_1,g_2,\ldots , g_N, E)$, for $ i=1,...,N$,  as functions of the couplings $g_k$,
of kinematical variables collectively denoted by $E$, and  $\hbar$,  taken to be zero for any classical theory. By taking the
expressions of these physical quantities  for some fixed
kinematics $\mu_j$ ($j=1,\ldots, N$)
\begin{eqnarray}
{  \sigma_1 (\mu_1)} & { =} & { f_1(\hbar, g_1,g_2,\ldots , g_N,\mu_1)\,,}\nonumber\\
{ \sigma_2(\mu_2)} & { =} & { f_2(\hbar, g_1,g_2,\ldots , g_N,\mu_2)\,,}\nonumber\\
         & { \cdots} &
\nonumber\\
{ \sigma_N (\mu_N)} & { =} & { f_N(\hbar, g_1,g_2,\ldots , g_N,\mu_N)\,,}\nonumber
\end{eqnarray}
we express the ${g_i}$ as functions of observables ${ \sigma_i(\mu_i)}$ 
\begin{eqnarray}
{ g_i} & { =} & { \phi_i(\hbar, \sigma_1(\mu_1),\sigma_2(\mu_2),\ldots , \sigma_N(\mu_N),\mu_1,\ldots , \mu_N)} ,\ i=1,\ldots, N .
\label{gis}
\end{eqnarray}
Next, we substitute the $g_i$ from Eq.~(\ref{gis}) in the calculated expressions of all physical quantities and obtain
\begin{eqnarray}
{ \sigma_{1}(E)} & { =} & { F_{1}(\hbar,\sigma_1(\mu_1),\sigma_2(\mu_2),\ldots , \sigma_N(\mu_N),\mu_1,\ldots , \mu_N, E)\,,}\nonumber\\
{ \sigma_{2}(E)} & { =} & { F_{2}(\hbar,\sigma_1(\mu_1),\sigma_2(\mu_2),\ldots , \sigma_N(\mu_N),\mu_1,\ldots , \mu_N, E)\,,}\nonumber\\
         & { \cdots} & \nonumber\\
{ \sigma_{N}(E)} & { =} & { F_{N}(\hbar,\sigma_1(\mu_1),\sigma_2(\mu_2),\ldots , \sigma_N(\mu_N),\mu_1,\ldots , \mu_N, E)\,,}\nonumber\\
{ \sigma_{N+1}(E)} & { =} & { F_{N+1}(\hbar,\sigma_1(\mu_1),\sigma_2(\mu_2),\ldots , \sigma_N(\mu_N),\mu_1,\ldots , \mu_N, E)\,,}\nonumber\\
         & { \cdots} &,
\nonumber
\end{eqnarray}
where the $F_i$ are some functions, whose specific form depends on the considered theory. In
this way we express all physical quantities in terms of the ``reference'' physical quantities taken at the normalization
points $\mu_i$, instead of the original ``bare'' parameters. Performed in the framework of QFT this procedure is known as
renormalization.  Notice that renormalization is non-perturbative if calculations are done non-perturbatively and perturbative
if we are using perturbation theory.

As a rule in QFTs, the  coefficients of the expansion of physical quantities in powers of the bare couplings $g_i$ are divergent.
In some theories divergences disappear after renormalization, that is, the expansion of all physical quantities in terms of
the ``reference'' physical quantities in the limit of a removed
regulator turn out to have finite coefficients to all orders. 
Such theories are called renormalizable (as e.g. the already mentioned case of QED).
To render the Green functions finite one often introduces renormalized fields and
corresponding field renormalization constants. However,  physical quantities do not depend on the choice of fields and,
therefore, one can also work with bare fields. To simplify the
discussion we formulate our considerations in
such a framework.  

In practice one uses  more conveniently chosen functions
$g_i^R(\mu) = {\cal G}_i(\sigma_1(\mu_1),\sigma_2(\mu_2),\ldots , \sigma_N(\mu_N),\mu_1,\ldots , \mu_N) $ ($j=1,\ldots, N$)
as new renormalized parameters, where $\mu$ stands collectively for all $\mu_1\,,\ldots , \mu_N$. 
Expressing the  original parameters $g_i$ in terms of the $g_i^R(\mu)$ 
\begin{eqnarray}
{g_i} & { =} & { \phi_i(\hbar, g^R_1(\mu),  g^R_2(\mu),\ldots , g^R_N(\mu),\mu_1,\ldots , \mu_N)} ,\ i=1,\ldots, N ,
\label{gisR}
\end{eqnarray}
and substituting in the Lagrangian of the given QFT,  we can express the Lagrangian in terms of the renormalized parameters. 
Usually, the functions ${\cal G}_i$ are chosen such that 
$${ \phi_i(0, g^R_1(\mu),  g^R_2(\mu),\ldots , g^R_N(\mu),\mu_1,\ldots , \mu_N)}=g_i^R(\mu) ,$$ 
leading to 
\begin{eqnarray}
  g_i & = & g_i^R(\mu)+\left[ { \phi_i(\hbar, g^R_1(\mu),  g^R_2(\mu),\ldots , g^R_N(\mu),\mu_1,\ldots , \mu_N)}
    -{ \phi_i(0, g^R_1(\mu),  g^R_2(\mu),\ldots , g^R_N(\mu),\mu_1,\ldots , \mu_N)} \right] \nonumber\\
 &=& g_i^R(\mu)+ \sum_{k=1}^\infty \hbar^k \delta g_i^k (g^R_1(\mu),  g^R_2(\mu),\ldots , g^R_N(\mu),\mu_1,\ldots , \mu_N), 
\label{gRandcts}
\end{eqnarray}
where the $\delta g_i^k$ are counterterms of $k$-th order in the loop
expansion which themselves can be expanded in powers 
of renormalized coupling constants.  Using the Lagrangian parameterized in terms of renormalized couplings and counterterms, we 
perform calculations of physical quantities in terms of the renormalized parameters $g_i^R(\mu)$. In renormalizable theories,
a perturbative calculation of physical quantities to any finite order in the renormalized coupling constants leads to 
Taylor series with finite coefficients. All counterterm contributions up to given order in the loop expansion and in the
coupling constants have to be taken into account in these calculations. If, for whatever reason, the expansion in any of these renormalized 
couplings needs to be summed  up to an infinite order, then this has
to be done systematically by also including the contributions stemming from
{\it all} counterterms. While such a calculation might be technically very complicated or even unfeasible,
the formalism of QFT renormalization defines very precisely and uniquely what has to be done.

From the modern point of view, renormalizable (in the traditional
sense) theories are viewed as leading order approximations to
effective field theories. The corresponding Lagrangians  contain an infinite number of terms compatible with underlying symmetries. 
A self-consistent EFT is renormalizable in the sense that all divergences appearing in physical quantities can be absorbed
in the redefinition of an infinite number of parameters of the effective Lagrangian (i.e.~$N$ is equal to infinity in an EFT). 
Despite the dependence on an infinite number of parameters,  EFTs do
not loose predictive power if  it is possible to
find renormalization schemes that ensure that low-energy observables
calculated with a specified accuracy depend only on a
finite number of {\it renormalized} parameters. Notice that physical quantities calculated to some finite order depend on
the choice of the renormalization scheme. While this dependence is
formally of higher order in the EFT expansion, the higher order corrections 
to physical quantities are suppressed only for appropriately chosen
renormalization conditions.
While it is always possible
to choose a scheme that destroys the validity of perturbation theory,
it is not always possible to find a scheme for which perturbative
series make sense. 
 A well-known example is QCD, where, on the one
hand, perturbation theory becomes very useful  at high energies, due to the asymptotic
vanishing of the renormalized coupling, provided  the renormalization
scale is chosen of the order of the
characteristic energy. 
On the other hand, it is impossible to find a renormalization scheme
at low energies which would lead to meaningful
perturbative expressions for physical quantities in powers of the renormalized coupling constant. 


\medskip

To be more specific, the Lagrangian of an EFT has the generic form
\begin{eqnarray}
{\cal L} =K(\psi)+ \sum_{i=1}^\infty g_i O_i(\psi)\,,
\label{EFTLagr}
\end{eqnarray}
where $K$ is the kinetic part, $O_i$ are interaction terms with bare couplings $g_i$ and $\psi$ stands collectively for
the fields.\footnote{It is understood that the considered EFT leads to a systematic perturbative expansion of physical quantities
for energies $E\ll Q$, where $Q$ is some large scale.} Because of the UV divergences one needs to introduce some kind of regularization.
The final expressions of physical quantities in terms of the renormalized 
couplings are finite in the limit of removed regularization and the results do not depend on the choice of 
the specific regularization scheme.  As mentioned above, to render the Green functions finite one often introduces renormalized
fields and corresponding field renormalization constants, however, 
physical quantities do not depend on the choice of fields and therefore one can also work with bare ones in the regularized theory.
To calculate physical quantities one expresses the bare couplings in terms of the renormalized ones as specified above and rewrites
the Lagrangian as
\begin{eqnarray}
  {\cal L} =K(\psi)+ \sum_{i=1}^\infty \left[g_i^R(\mu)+ \sum_{k=1}^\infty \hbar^k \delta g_i^k (g^R_1(\mu),  g^R_2(\mu),\ldots ,
    g^R_N(\mu),\mu_1,\ldots , \mu_N, \Lambda)\right] O_i(\psi)\,,
\label{EFTLagrR}
\end{eqnarray}
where we also indicate the dependence of the counterterms on the regularization scheme by including the parameter $\Lambda$ in the
list of arguments.  While one obviously has a huge freedom of choosing
the renormalization conditions,
performing consistent calculations of physical quantities up to any
order (either finite or infinite) in any coupling $g_j^R$ for any
particular choice requires the inclusion of {\it all} contributions up
to a given order, generated by the Lagrangian of Eq.~(\ref{EFTLagrR}).
It is 
difficult to implement such a formally consistent program in the few-body sector of the low-energy EFT of the strong interactions,
where one needs to sum up an infinite number of diagrams. 
The problem is caused by the necessity of taking into account contributions of an infinite number of counterterms with 
growing complexity of the operator structures $O_i(\psi)$ in the (heavy baryon) formulation of the EFT.

The best available approach to the non-relativistic formulation of an EFT with pions and nucleons as explicit degrees of freedom
is the cutoff EFT which relies on the inclusion of a finite number of counterterms and keeping the cutoff parameter
such that the difference between the properly renormalized and
actually obtained results is of a higher order than the
accuracy of the current calculation
\cite{Lepage:1997cs,Lepage:1999kt,Gegelia:1998iu,Gegelia:2004pz,Epelbaum:2006pt,Epelbaum:2008ga,Epelbaum:2019kcf}.

A convenient way of implementing the quantum field theoretical renormalization is to apply the BPHZ subtractive procedure using
Zimmerman's forest formula, see, e.g., Ref.~\cite{Collins:1984xc}. This subtractive approach is equivalent to the above outlined 
renormalization by including the contributions of all counterterms appearing at a given order. Within the subtractive approach,
instead of explicitly including the  counterterm contributions, one renormalizes the loop diagrams by subtracting 
sub-divergences and overall divergences using some fixed renormalization conditions and 
treats all coupling constants as renormalized finite parameters.  In the next section we consider a particular subtractive
quantum field theoretical renormalization for the non-perturbative solutions to integral equations with singular
effective potentials which removes all divergences from all diagrams generated by iterations of the integral equations.

\section{\label{contactintpotential}Integral equation  with a singular potential and renormalization}

In this section we consider one particular realization of a subtractive EFT renormalization for solutions of integral
equations with singular  potentials.  Our scheme is a generalization of the one employed in Refs.~\cite{Weinberg:rz,Gegelia:1998gn}
to the case of the one-pion-exchange potential and differs from those of 
Refs.~\cite{Frederico:1999ps,Timoteo:2005ia,Timoteo:2010mm,Yang:2007hb,Yang:2009kx}. In particular, Refs.~\cite{Yang:2007hb,Yang:2009kx} 
apply a subtraction scheme at vanishing energy, whose realization in terms of local counterterms of the effective Lagrangian
is unclear to us. On the other hand Refs.~\cite{Frederico:1999ps,Timoteo:2005ia,Timoteo:2010mm}
demand that the  scattering amplitude calculated at a given order must be independent of the subtraction point. As emphasized
in the previous section, our understanding is that perturbative
amplitudes calculated up to any finite order are
renormalization-scheme  dependent.  A toy-model example considered in the next section demonstrates how problematic it may turn out to find a
renormalization scheme, which is not only conceptually consistent but also useful in describing the data. 

We emphasize that there is a conceptual difference between the EFT renormalization of the scattering
amplitude obtained by solving
the Lippmann-Schwinger (LS) equation with singular potentials, that takes into account contributions
of an infinite number  of counterterms with operator structures of increasing complexity in
each partial wave, and self-adjoint extensions of singular Hamiltonians in the framework of
quantum mechanics, that ensure finiteness of the calculated observables in the removed regularization
limit \cite{Frank:1971xx}.

\medskip

We start by considering the LS integral equation for an off-shell scattering amplitude which we write symbolically as
\begin{equation}
T=V+\hbar \,V GT~.
\label{one}
\end{equation}
To avoid any non-analytic dependence on the quark masses of the counterterms in the few-body sector of the chiral EFT,
we split the potential into a singular $V_S$ and a regular $V_R$ part,
\begin{equation}
V= V_S+V_R\,,
\label{pot}
\end{equation}
where $V_S$ has only a polynomial dependence on the quark masses and $V_R$ is such that its overlap in loop diagrams with $V_S$,
i.e. $V_S G V_R$, is finite.  For the potential of Eq.~(\ref{pot}) 
the solution to Eq.~(\ref{one}) can be written in the form
\begin{equation}
T=T_S+(1+\hbar \,T_S\,G)\,T_C (1+\hbar \,G\,T_S)~,
\label{Td}
\end{equation}
where $T_S$ and $T_C$ satisfy the equations
\begin{equation}
T_S=V_S+\hbar \,V_S\,G\,T_S,
\label{OEQ1S}
\end{equation}
and 
\begin{equation}
T_C=V_R+\hbar \,V_R\,G\,(1+\hbar \,T_S G)\,T_C\,.
\label{OEQ13}
\end{equation}

To carry out subtractive renormalization, we take into account contributions of counter terms such that  the
amplitude $T_S$ in Eq.~(\ref{Td}) is replaced by the renormalized amplitude $T_S^r$. 
In particular, we consider iterations of Eq.~(\ref{OEQ1S}) and perform BPHZ subtractions in each of
an infinite number of terms such that the renormalized series has the form: 
\begin{equation}
T_S^r=V_S+ \hbar \, V_S\,(G-G_e)\, V_S+\hbar^2 \,V_S\,(G-G_e)\, V_S \,(G-G_e)\, V_S + \cdots .
\label{OEQ1It}
\end{equation}
In the first iteration, $\hbar \,V_S\,(G-G_e)\, V_S$, the result of the original one-loop diagram
$\hbar \, V_S\,G\, V_S$ is
expanded around $E=-E_\mu$ and the first $N+1$ terms are subtracted by adding $\hbar \,V_S\,(-G_e)\, V_S$, where 
\begin{equation}
G_e=
\sum_{i=0}^N \frac{1}{i!} (E+E_\mu)^i \frac{d^i G(E)}{(d E
  )^i}\bigg|_{E=-E_\mu}\,.  
\label{Gs}
\end{equation} 
That is, we expand $G\equiv G(E)$ at a fixed value $E=-E_\mu$ and subtract
first several terms, the number of the subtracted terms depending on the UV behaviour of the  singular potential.
The $N+1$ terms, subtracted at one-loop order, if expanded in powers of 
momenta and energy, correspond to an infinite number of local counterterms in the effective Lagrangian.
Notice here that from the point of view of UV divergences, it would be sufficient to subtract
only first several terms in the expansion in powers of momenta and energy.
Thus our choice  of the renormalization scheme corresponds to finite over-subtractions. This applies to
all orders in the loop expansion.
In the second iteration in Eq.~(\ref{OEQ1It}), two one-loop 
sub-divergences are subtracted from the original two-loop diagram
$\hbar^2 \,V_S\, G\, V_S \, G\, V_S$
by adding $\hbar^2 \,V_S\,(-G_e)\, V_S \,G\, V_S$ and $\hbar^2 \,V_S\,G\, V_S \,(-G_e)\, V_S$, and
then the overall divergence is subtracted by adding $\hbar^2 \, V_S\,G_e\, V_S \,G_e\, V_S$. 
Again, these overall two-loop subtractions correspond to taking into account contributions of an infinite
number of counter-terms.
Renormalization works analogously for
further iterations. 
The renormalized amplitude $T_S^r$ satisfies the equation 
\begin{equation}
T_S^r=V_S+\hbar \,V_S\,(G-G_e)\,T_S^r~.
\label{OEQ1}
\end{equation}
The reason why we over-subtract each term in series of Eq.~(\ref{OEQ1It}) is that only this way we can
obtain the full renormalized amplitude in closed form 
as a solution to an integrals equation.

The effective potential $V_e$ which includes all necessary counterterms subtracting sub-divergences and
overall divergences
in all iterations of Eq.~(\ref{OEQ1}) can be obtained by solving the following equation 
\begin{equation}
V_e=V_S-\hbar\, V_S G_eV_e\,.
\label{EQGa}
\end{equation}
That is, the subtracted amplitude $T_S^r$ can also be obtained by solving the equation 
\begin{equation}
T_S^r=V_e+\hbar \,V_e\,G\,T_S^r\,,
\label{OEQ1R}
\end{equation}
where $V_e$ is obtained by solving Eq.~(\ref{EQGa}).

\medskip

The final renormalized expression of the amplitude has the form
\begin{equation}
T^r=T_S^r+(1+\hbar\,T_S^r\,G)\,T_C^r (1+\hbar\,G\,T_S^r),
\label{TdR}
\end{equation}
where  $T_C^r$ satisfies the equation
\begin{equation}
T_C^r=V_R+\hbar\,V_R\,G\,(1+\hbar\,T_S^r G)\,T_C^r.
\label{OEQ13R}
\end{equation}
For practical applications it is convenient to obtain the final renormalized amplitude $T^r$ in a
single step as a solution to an integral equation. Below we show that indeed $T^r$ can be
obtained by solving the following integral equation
\begin{equation}
T^r=V_S+\tilde V_R +\hbar\,(\tilde V_S+\tilde V_R)\,G\,T^r,
\label{OEQ1P}
\end{equation}
where $\tilde V_S= V_S (G-G_e)/G $ and $\tilde V_R=V_R+ \hbar\,(V_S-\tilde V_S)G V_R$.\footnote{Eq.~(\ref{OEQ1P})
  is convenient because it contains only convergent integrals.}
To this end we write the solution to Eq.~(\ref{OEQ1P}) as $T_S^r+T_1$ and show that $T_1$ is identical to
the second term in the right hand side of Eq.~(\ref{TdR}). 
Indeed, substituting $T_S^r+T_1$ in  Eq.~(\ref{OEQ1P}) we obtain the following equation for $T_1$
\begin{equation}
T_1=\tilde V_R(1+\hbar\,G T_S^r) +\hbar\,(\tilde V_S+\tilde V_R)\,G\,T_1\,.
\label{OEQT1}
\end{equation}
Next, we define the amplitude  $T_x$ via 
\begin{equation}
T_1=:  (1+ \hbar\,T_S^r G)T_x (1+\hbar\,G T_S^r)
\label{TxDef}
\end{equation}
and obtain from Eq.~(\ref{OEQT1})
\begin{equation}
(1-\hbar\,\tilde V_S G+ \hbar\,(1-\hbar\,\tilde V_S G)T_S^r G)T_x=\tilde V_R +\hbar\,\tilde V_R \,G\,
(1+ \hbar\,T_S^r G)T_x\,.
\label{OEQTx}
\end{equation}
Writing the equation (\ref{OEQ1}) in the form 
\begin{equation}
(1-\hbar\,\tilde V_s G)T_S=V_S,
\label{Tsin}
\end{equation}
and using it in Eq.~(\ref{OEQTx}), we obtain
\begin{equation}
(1- \hbar\,\tilde V_S G+ \hbar\, V_S G)T_x=\tilde V_R +\hbar\,\tilde V_R \,G\,(1+ \hbar\,T_S^r G)T_x\,.
\label{OEQTx2}
\end{equation}
Using the definition of $\tilde V_R$, Eq.~(\ref{OEQTx2}) reduces to 
\begin{equation}
T_x = V_R + \hbar\,V_R \,G\,(1+ \hbar\,T_S^r G)T_x\,.
\label{OEQTx3}
\end{equation}
Comparing Eqs.~(\ref{OEQTx3}) and  (\ref{OEQ13R}), we see that $T_x$ is identical with $T_C^r$ and hence,
as evidenced from Eq.~(\ref{TxDef}),  $T_1$ is identical to the second term in the right hand side of
Eq.~(\ref{TdR}). This proves that indeed $T^r$ satisfies Eq.~(\ref{OEQ1P}). 

We emphasize that we consider the above subtractive scheme not because it is 
amongst the best ones for, e.g., nucleon-nucleon scattering in
chiral EFT,  but rather since it can be easily applied in practice
and, therefore, well suited for demonstration purposes. 
As noticed in Refs.~\cite{Gegelia:1999ja,Epelbaum:2017byx}  in the nucleon-nucleon scattering problem, it is preferable to use
subtraction schemes with two scales
corresponding to the leading- and higher-order potentials.

\section{Application to a  toy model}
\label{OPE}

As a specific toy-model example of how to apply the discussed renormalization scheme, we construct
an effective potential for the underlying ``fundamental'' potential
considered in Ref.~\cite{Epelbaum:2018zli}
\begin{eqnarray}
V(r) &=& \frac{\alpha  \left(e^{- m_1 r}-e^{-M
   r}\right)}{r^3}+\frac{\alpha  \left(m_1-M\right) e^{- m_1 r}}{r^2}
   +\frac{\alpha  \left(M-m_1\right){}^2 e^{-m_2 r}}{2 r} \nonumber\\
   &-& \frac{1}{6} \alpha  \left(2 m_1-3
   m_2+M\right) \left(M-m_1\right){}^2 e^{- m_1 r}~,
\label{potentialdef}
\end{eqnarray}
where $M$ is the light mass and  the heavy masses $m_1$, $m_2$ represent the large scales. 
Our choice of parameters is $\alpha=-36\, {\rm GeV^{-2}}$, $M=0.1385$~GeV, $m_1=0.75$~GeV and 
$m_2=1.15$~GeV. The strength of the interaction $\alpha$ is 
taken equal for all terms, so that
the potential $V(r)$ vanishes for $r\to 0$ and it behaves as $-\alpha\, e^{-M r}/r^3$ for large $r$.
More details on this model can be found in Ref.~\cite{Epelbaum:2018zli}.

We consider the LS equation for the S-wave scattering amplitude in the center-of-mass frame of two particles 
with unit masses 
\begin{equation}
t_E(p,p\,') =  v (p,p\,')+  \int_0^\infty\frac{d q\, q^2}{2\pi^2} \,v(p,q)\,\frac{1}{E-q^2+i\,\epsilon}\,t_E(q,p'),
\label{tEq}
\end{equation}
with $E=k^2/1 \, {\rm GeV}$ being the total energy, and $p=|{\bf p}|$, $p'=|{\bf p}' | $  the relative momenta of 
the incoming and  outgoing particles, respectively.  We set  $\hbar=1$ in these calculations. 

It is easy to construct the corresponding EFT for the underlying ``fundamental'' potential in Eq.~(\ref{potentialdef}).
The LO EFT interaction consists of a delta 
potential and the long-range part $ -{\alpha\, e^{-M r}}/{r^3}$, which is 
singular if extended to the small-$r$ region.
Following Ref.~\cite{Epelbaum:2018zli}, we choose the coupling constant $\alpha=-36 \ {\rm GeV^{-2}}\approx -1/(0.167 \ {\rm GeV})^2$ 
such that the full LO potential is non-perturbative for 
the momenta $k\sim M=0.1385 \ {\rm GeV}$. 
 A simple UV analysis shows that the LO potential is perturbatively non-renormalizable, i.e. 
to remove the divergences from its iterations one needs to introduce counterterms of higher orders (in momenta and energy) 
which themselves generate new divergences, etc., up to infinity. 

\begin{figure}[t]
\includegraphics[width=\textwidth]{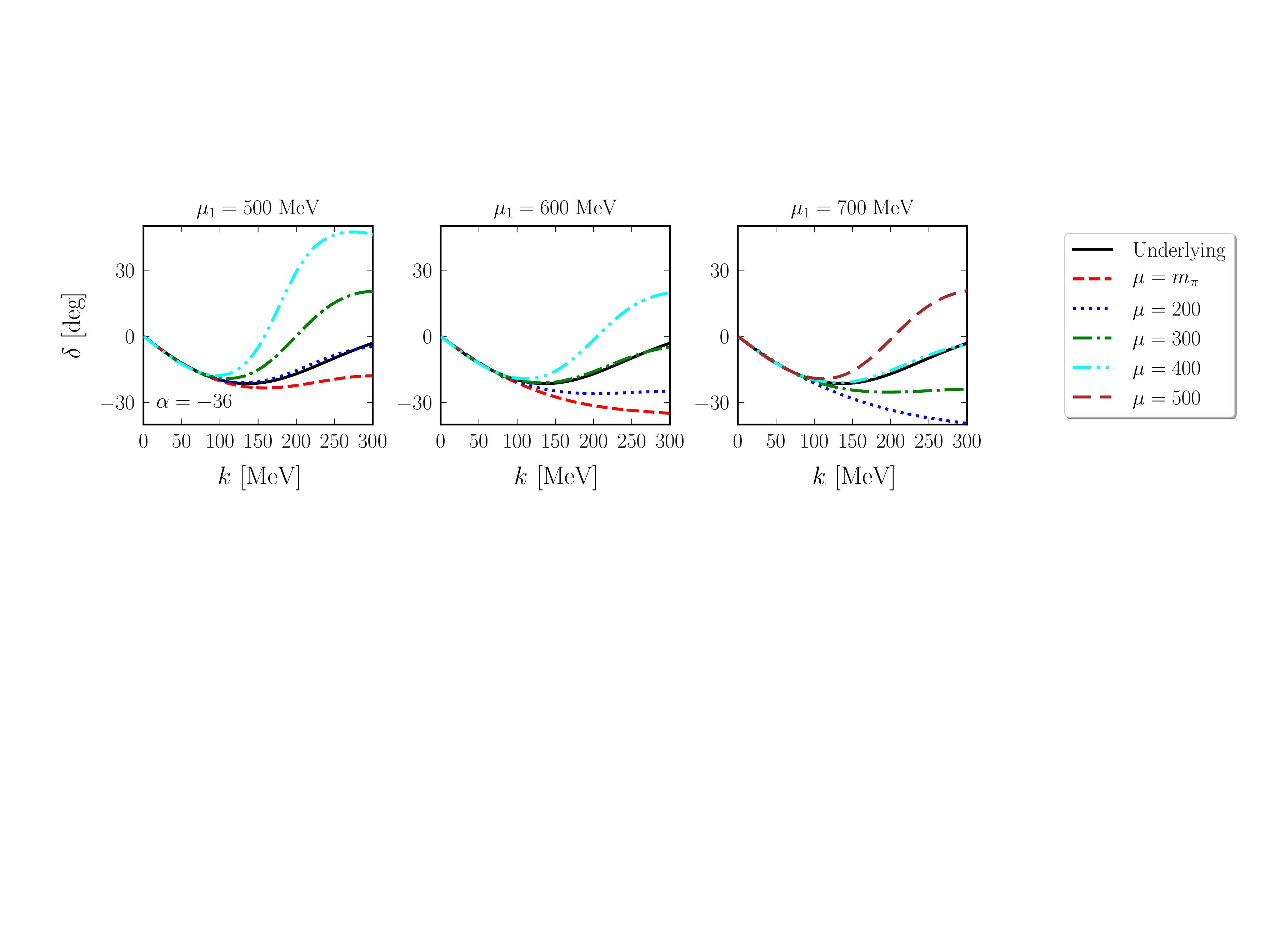}
\caption{The leading-order S-wave phase shift as a function of the
  center-of-mass momentum $k$. 
Solid
  (black) lines correspond to the underlying toy model while 
  dashed lines show the LO phase shifts for various values of $\mu$
  and $\mu_1$. For each choice of the scale $\mu_1$, four
  representative choices for the scale $\mu$ are shown to
  demonstrate the trends.}
\label{Pots} 
\end{figure}

We apply the renormalization scheme described in the previous section to the LO effective potential, for
which we take
\begin{eqnarray}
V_S(r) &=& c \, \delta^{(3)}(\vec r)  
-\frac{\alpha\,  e^{- \mu_1 r}}{r^3}+
\frac{\alpha  \left(M-\mu _1\right) e^{- \mu _1 r}}{r^2}-\frac{\alpha 
   \left(M-\mu _1\right){}^2 e^{-\mu _1 r}}{2 r} , \nonumber\\
V_R(r)   &=& -\frac{\alpha  \left(e^{-M r}-e^{-\mu _1 r}\right)}{r^3}-
\frac{\alpha \left(M-\mu _1\right) e^{-\mu _1 r}}{r^2}+\frac{\alpha 
   \left(M-\mu _1\right){}^2 e^{-\mu _1 r}}{2 r} \nonumber\\
   &=& \frac{1}{6} \alpha  \left(M-\mu _1\right){}^3-\frac{r}{24} \, \alpha  
   \left(M-\mu _1\right){}^3 \left(3 \mu _1+M\right) + {\cal O}(r^2)\, ,
\label{LOEFTpotentialdef}
\end{eqnarray}
where $\mu_1$ is the scale parameterizing the division of the potential into the singular
and regular parts. To obtain the singular part, we expanded the LO potential in a Taylor series of
the mass $M$ around $M=\mu_1$  and kept the first three terms in the power series of $M-\mu_1$. Notice
that since our toy model is not generated by a
chirally invariant Lagrangian, we obtain a power series expansion in
$M$ and not in $M^2$. 
As specified in section~\ref{contactintpotential}, the number of
subtracted terms when defining the regular potential $V_R$ is dictated
by the condition that all diagrams generated by iterating
Eq.~(\ref{OEQ1P}) remain finite. 
We fix the renormalized coupling $c_R$ by the requirement to reproduce the low-energy phase shift
of the underlying toy-model potential at a fixed value of the on-shell
momentum $k$. 
Subtractions of loop integrals are performed at $E=-\mu^2/(1\, {\rm GeV})$.

Our results for the phase shift for various choices of the renormalization scales are shown in
Fig.~\ref{Pots} together with the phase shifts of the underlying
toy-model to be regarded as synthetic data. The scale $\mu_1$ that parametrizes
the ambiguity in the definition of the regular part of the potential
should be chosen  of the order of the hard scale in
the problem. Using too soft values of the scale $\mu_1$ is, in general, expected to
reduce the applicability range of the EFT.  
As seen from Fig.~\ref{Pots}, one can find a range of values of the
subtraction scales $\mu$ and $\mu_1$, for which
the description of the synthetic data is reasonably good for the LO
approximation. However, the 
results for the momenta $k \gtrsim  M$ are sensitive to the changes of these values. As mentioned in
section~\ref{ren}, approximate perturbative expressions in QFTs always depend on the choice of the
renormalization scheme.

\begin{figure}[t]
\includegraphics[width=0.85\textwidth]{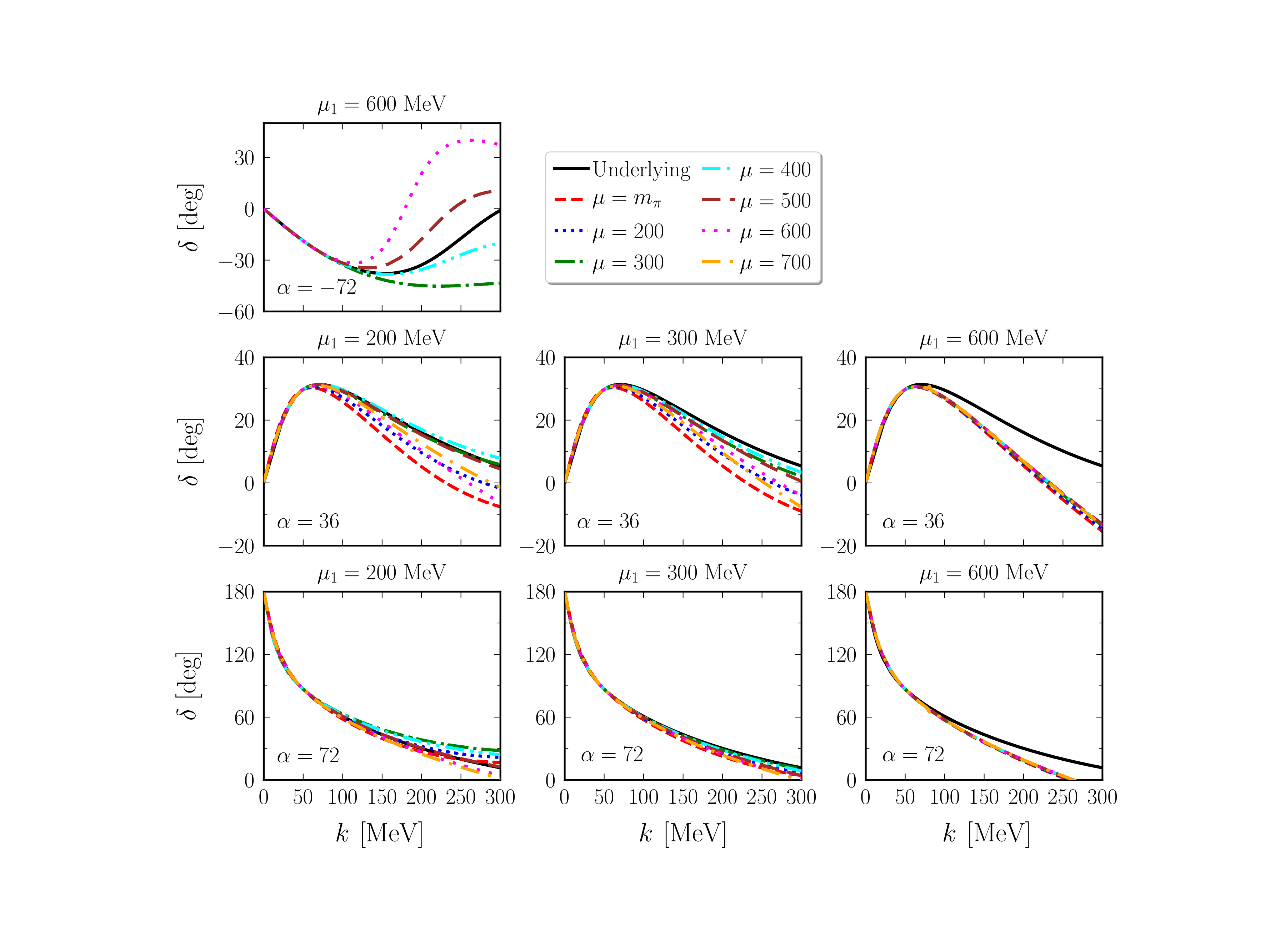}
\caption{The leading-order S-wave phase shifts as functions of the
  center-of-mass momentum $k$  for different values of the coupling $\alpha$. Solid
  (black) lines correspond to the underlying toy model while 
  dashed lines show the LO phase shifts for various values of $\mu$ and $\mu_1$.}
\label{Pots2} 
\end{figure}

To demonstrate that the observed reasonable description of the
synthetic data within the employed subtractive renormalization scheme
is not just an artifact of the specific choice  of the coupling
constant $\alpha$ in the underlying toy model, we also consider
the cases corresponding to different values of $\alpha$. Specifically,
by changing the sign of the coupling to $\alpha = 36$, we arrive at a
toy model with the attractive long-range potential. We also increase
the strength of the long-range potential by a factor of two and
consider the cases of $\alpha = \pm 72$.  Our results for these three
additional models  are visualized in Fig.~\ref{Pots2} for a number of 
representative choices of the scales $\mu$ and $\mu_1$. For a strong 
repulsive long-range interaction corresponding to $\alpha = -72$, we
only show the results for $\mu_1 = 600$~MeV since the observed pattern 
is analogous to the already considered toy-model with $\alpha = -36$.   
Phase shifts corresponding to the attractive long-range potential are plotted in
the second and third rows of Fig.~\ref{Pots2}. The behavior of the
phase shifts indicates the appearance of a low-lying virtual (real) bound state
for $\alpha = 36$ ($\alpha = 72$). Compared to the repulsive case, we
find an even larger range of subtraction scales leading to a
reasonable description of the phase shifts. In these particular cases,
softening the scale $\mu_1$ effectively amounts to a resummation of
the subleading contact interaction which helps to improve the
reproduction of the effective range.\footnote{Notice that in the realistic case of neutron-proton
scattering in the $^1$S$_0$ channel, the subleading contact
interaction was argued to require a nonperturbative treatment in Ref.~\cite{Epelbaum:2015sha}.} 
Still, choosing $\mu_1$ of the order of the hard scale such as e.g.~$\mu_1 =
600$~MeV leads to an adequate LO description of the synthetic data.
Furthermore, fully in line with the findings of Ref.~\cite{Epelbaum:2015sha},
the scale $\mu$ can be varied in a broad range and set to soft values
since the systems under consideration are not too close to the unitary limit.

\section{Summary}
\label{summary}

In this paper we discussed the problem of renormalization of  singular potentials  in the low-energy effective field theory of the 
strong interactions.  We outlined the standard procedure of quantum field theoretical renormalization emphasizing its general definition
that applies  to both non-perturbative as well as perturbative
calculations. The main message of our work is that chiral EFT in the
few-nucleon sector does not require the
invention of any new non-perturbative renormalization
 that would go beyond what is already well-established in
quantum field theory.

To demonstrate that the problem lies not in conceptual issues but
rather in the complicated technical implementation of the
standard renormalization approach,  we
proposed a way of implementing the
subtractive renormalization for solutions of integral
equations with singular potentials.  The suggested renormalization scheme corresponds to subtracting all divergences appearing
at a given order, and also (over-)subtracting finite pieces with some fixed normalization conditions. 
While being just one of infinitely many possibilities of fixing the normalization
conditions, and very likely not the best one, the suggested scheme has
the advantage of being easily applicable in practice
and serves well our purposes of demonstrating a self-consistent EFT renormalization. 

As an example to demonstrate these rather general statements, we applied the considered subtractive renormalization scheme to the
toy model of Ref.~\cite{Epelbaum:2018zli}. While there exists a region of values of the renormalization scales for which the
description of the ``data'' is good for the LO approximation, the 
results are sensitive to the changes of these values as one expects on
general grounds.

\section*{Acknowledgments}

This work was supported in part by BMBF (Grant No. 05P18PCFP1), by DFG (Grant No. 426661267), by DFG and NSFC through funds provided to the
Sino-German CRC 110 ``Symmetries and the Emergence of Structure in QCD" (NSFC
Grant No.~11621131001, DFG Grant No.~TRR110), by the 
Georgian Shota Rustaveli National
Science Foundation (Grant No. FR17-354), by VolkswagenStiftung (Grant no. 93562)
 and by the CAS President's International
Fellowship Initiative (PIFI) (Grant No.~2018DM0034).


\begin{thebibliography}{1}

\bibitem{Weinberg:rz}
S.~Weinberg,
Phys.\ Lett.\ B {\bf 251}, 288 (1990).

\bibitem{Bedaque:2002mn}
  P.~F.~Bedaque and U.~van Kolck,
  Ann.\ Rev.\ Nucl.\ Part.\ Sci.\  {\bf 52}, 339 (2002),
  [nucl-th/0203055].

\bibitem{Epelbaum:2005pn}
  E.~Epelbaum,
  Prog.\ Part.\ Nucl.\ Phys.\  {\bf 57}, 654 (2006),
  [arXiv:nucl-th/0509032].

\bibitem{Epelbaum:2008ga}
  E.~Epelbaum, H.~-W.~Hammer and U.-G.~Mei\ss ner,
  Rev.\ Mod.\ Phys.\  {\bf 81}, 1773 (2009),
  [arXiv:0811.1338 [nucl-th]].
  
\bibitem{Birse:2009my}
  M.~C.~Birse,
  PoS CD {\bf 09}, 078 (2009)
  [arXiv:0909.4641 [nucl-th]].

\bibitem{Machleidt:2011zz}
  R.~Machleidt and D.~R.~Entem,
  Phys.\ Rept.\  {\bf 503}, 1 (2011),
  [arXiv:1105.2919 [nucl-th]].
  
\bibitem{Epelbaum:2012vx}
  E.~Epelbaum and U.-G.~Mei{\ss}ner,
  Ann.\ Rev.\ Nucl.\ Part.\ Sci.\  {\bf 62}, 159 (2012),
  [arXiv:1201.2136 [nucl-th]].

\bibitem{Valderrama:2016koj}
  M.~P.~Valderrama,
  Int.\ J.\ Mod.\ Phys.\ E {\bf 25}, 1641007 (2016),
  [arXiv:1604.01332 [nucl-th]].

\bibitem{Hammer:2019poc} 
  H.-W.~Hammer, S.~K\"onig and U.~van Kolck,
  arXiv:1906.12122 [nucl-th].

\bibitem{Machleidt:2020vzm}
  R.~Machleidt and F.~Sammarruca,
  arXiv:2001.05615 [nucl-th].

\bibitem{Weinberg:mt}
S.~Weinberg, ``The Quantum Theory Of Fields. Vol. 1,2:
Foundations, Modern applications'' {\it  Cambridge, UK: Univ. Pr. (1995)}.

\bibitem{Epelbaum:2018zli} 
  E.~Epelbaum, A.~M.~Gasparyan, J.~Gegelia and U.-G.~Mei\ss ner,
  Eur.\ Phys.\ J.\ A {\bf 54}, 186 (2018),
  [arXiv:1810.02646 [nucl-th]].
  
\bibitem{Lepage:1997cs}
G.~P.~Lepage,
arXiv:nucl-th/9706029.

\bibitem{vanKolck:2020llt} 
  U.~van Kolck,
  arXiv:2003.06721 [nucl-th].

\bibitem{Beane:1997pk} 
  S.~R.~Beane, T.~D.~Cohen and D.~R.~Phillips,
  Nucl.\ Phys.\ A {\bf 632}, 445 (1998),
  [nucl-th/9709062].

\bibitem{Lepage:1999kt}
G.~P.~Lepage, {\it Conference summary,}
{Prepared for INT Workshop on Nuclear Physics with EFT, Seattle,
Washington, 25-26 Feb 1999.}
  
  
\bibitem{Gegelia:1998iu}
  J.~Gegelia,
  J.\ Phys.\ G {\bf 25}, 1681 (1999),
  [nucl-th/9805008].

  
\bibitem{Gegelia:2004pz}
  J.~Gegelia and S.~Scherer,
  Int.\ J.\ Mod.\ Phys.\ A {\bf 21}, 1079 (2006),
  [nucl-th/0403052].

\bibitem{Epelbaum:2006pt}
  E.~Epelbaum and U.-G.~Mei\ss ner,
  Few Body Syst.\  {\bf 54}, 2175 (2013),
  [nucl-th/0609037].

\bibitem{Epelbaum:2019kcf} 
  E.~Epelbaum, H.~Krebs and P.~Reinert,
  arXiv:1911.11875 [nucl-th].


\bibitem{Collins:1984xc}
  J.~C.~Collins,
  ``{\it Renormalization. An introduction to renormalization, the renormalization group, and the operator product expansion},''
{\it  Cambridge, Uk: Univ. Pr. ( 1984) 380p}.

  
\bibitem{Gegelia:1998gn} 
  J.~Gegelia,
  Phys.\ Lett.\ B {\bf 429}, 227 (1998).

  
\bibitem{Frederico:1999ps} 
  T.~Frederico, V.~S.~Timoteo and L.~Tomio,
  Nucl.\ Phys.\ A {\bf 653}, 209 (1999),
  [nucl-th/9902052].
  
\bibitem{Timoteo:2005ia} 
  V.~S.~Timoteo, T.~Frederico, A.~Delfino and L.~Tomio,
  Phys.\ Lett.\ B {\bf 621}, 109 (2005),
  [nucl-th/0508006].

\bibitem{Timoteo:2010mm} 
  V.~S.~Timoteo, T.~Frederico, A.~Delfino and L.~Tomio,
  Phys.\ Rev.\ C {\bf 83}, 064005 (2011),
  [arXiv:1006.1942 [nucl-th]].


\bibitem{Yang:2007hb} 
  C.-J.~Yang, C.~Elster and D.~R.~Phillips,
  Phys.\ Rev.\ C {\bf 77}, 014002 (2008),
  [arXiv:0706.1242 [nucl-th]].


\bibitem{Yang:2009kx} 
  C.~J.~Yang, C.~Elster and D.~R.~Phillips,
  Phys.\ Rev.\ C {\bf 80}, 034002 (2009),
  [arXiv:0901.2663 [nucl-th]].
 
\bibitem{Frank:1971xx} 
  W.~Frank, D.~J.~Land and R.~M.~Spector,
  Rev.\ Mod.\ Phys.\  {\bf 43}, 36 (1971).
  doi:10.1103/RevModPhys.43.36

  
\bibitem{Gegelia:1999ja} 
  J.~Gegelia,
  Phys.\ Lett.\ B {\bf 463}, 133 (1999),
  [nucl-th/9908055].

\bibitem{Epelbaum:2017byx} 
  E.~Epelbaum, J.~Gegelia and U.-G.~Mei\ss ner,
  Nucl.\ Phys.\ B {\bf 925}, 161 (2017),
  [arXiv:1705.02524 [nucl-th]].

\bibitem{Epelbaum:2015sha} 
  E.~Epelbaum, A.~M.~Gasparyan, J.~Gegelia and H.~Krebs,
  Eur.\ Phys.\ J.\ A {\bf 51}, 71 (2015),
  [arXiv:1501.01191 [nucl-th]].
\end{thebibliography}
\end{document}